\documentclass[%
 reprint,
nofootinbib,
 amsmath,amssymb,
 bibtex,
 aps,
]{revtex4-2}

\usepackage{xcolor}

\usepackage[normalem]{ulem}
\newcommand{\dd}{\text{d}}

\usepackage[export]{adjustbox}

\usepackage{graphicx}% Include figure files
\usepackage{dcolumn}% Align table columns on decimal point
\usepackage{bm}% bold math
\usepackage{amsmath, amssymb}
\usepackage{graphicx}
\usepackage{multirow}

\usepackage{tikz}
\usepackage{tkz-euclide}
\usetikzlibrary{shapes.misc}
\usetikzlibrary{decorations.pathmorphing}	% To create gluons and vector bosons
\tikzset{
    v/.style={decorate, decoration={snake, segment length=3mm, amplitude=0.75mm}, draw},
    f/.style={draw=black, postaction={decorate},
        decoration={markings,mark=at position .6 with {\arrow[very thick]{latex}}}},
    fb/.style={draw=black, postaction={decorate},
        decoration={markings,mark=at position .4 with {\arrowreversed[very thick]{latex}}}},
    fnar/.style={draw=black},
    g/.style={decorate, draw=black,
        decoration={coil,amplitude=3pt, segment length=3.5pt}},
    s/.style={dashed,draw=black, postaction={decorate},
        decoration={markings,mark=at position .55 with {\arrow[very thick]{latex}}}},
    sb/.style={dashed,draw=black, postaction={decorate},
        decoration={markings,mark=at position .55 with {\arrowreversed[draw=black,very thick]{latex}}}},
    snar/.style={dashed,draw=black,line width =1.25pt},
    cross/.style={cross out, draw=black, minimum size=2*(#1-\pgflinewidth), inner sep=0pt, outer sep=0pt},
cross/.default={3pt},
}

\def\be{\begin{equation}}
\def\ee{\end{equation}}

\newcommand{\braket}[1]{\left\langle #1 \right\rangle}
\newcommand{\sbraket}[1]{\left[ #1 \right]}
\newcommand{\asbraket}[3]{\left\langle #1 \vert #2 \vert #3 \right]}
\newcommand{\sabraket}[3]{\left[ #1 | #2 | #3 \right\rangle}
\newcommand{\bra}[1]{ \left\langle #1 \right|}
\newcommand{\sbra}[1]{ \left[ #1 \right|}
\newcommand{\ket}[1]{ \left| #1 \right\rangle}
\newcommand{\sket}[1]{ \left| #1 \right]}

\newcommand{\rket}[1]{ \left| #1 \right)}
\newcommand{\rbraket}[1]{\left( #1 \right)}
\newcommand{\rrbraket}[3]{\left( #1 \vert #2 \vert #3 \right)}

\newcommand{\al}[1]{\begin{align}\begin{aligned} #1 \end{aligned}\end{align}}

\begin{document}

\preprint{APS/123-QED}

\title{Characterising Dark Matter-induced neutrino potentials}% Force line breaks with \\
 \author{Gabriel M. Salla}%
 \email{gabriel.massoni.salla@usp.br}
%  \altaffiliation[Also at ]{Physics Department, XYZ University.}%Lines break automatically or can be forced with \\

\affiliation{%
 Instituto de F\'isica, Universidade de S\~ao Paulo, C.P. 66.318, 05315-970 S\~ao Paulo, Brazil
}%

% \collaboration{MUSO Collaboration}%\noaffiliation

%  \homepage{http://www.Second.institution.edu/~Charlie.Author}
% \affiliation{
%  Second institution and/or address\\
%  This line break forced% with \\
% }%
% \affiliation{
%  Third institution, the second for Charlie Author
% \affiliation{%
%  Authors' institution and/or address\\
%  This line break forced with \textbackslash\textbackslash
% }%

% \collaboration{CLEO Collaboration}%\noaffiliation

\date{\today}% It is always \today, today,

\begin{abstract}
% Abstract must have less than 600 characters. Letter must have less than 3,750 words.\\
% \begin{description}
% Secondary publications and information retrieval purposes.
% You may use the \texttt{description} environment to structure your abstract;
% use the optional argument of the \verb+\item+ command to give the category of each item. 
% \end
In this paper we explore interactions between neutrinos and Dark Matter. In particular, we study how the propagation of astrophysical neutrinos can be modified by computing the most general potential generated by the galactic DM background. We use on-shell techniques to compute this potential in a completely model independent way and obtain an expression valid for any Dark Matter mass and spin. Afterwards, we use this expression to analyse under what circumstances such potential can be important at the phenomenological level, and we find that under some assumptions only ultra light scalar Dark Matter could be of any relevance to oscillation experiments.
\end{abstract}

\maketitle

%\tableofcontents

%\section{\label{sec:intro}Introduction}

\textit{Introduction} - Since the discovery of astrophysical neutrinos by IceCube \cite{1304.5356}, it became crucial to comprehend how neutrinos propagate through the cosmos. In particular, the large amount of Dark Matter (DM) present in the path of these high-energetic neutrinos raises the question whether or not an interaction with the galactic DM background could be phenomenologically relevant. Previous studies have explored the phenomenology of many models with neutrino-DM ($\nu$DM) interactions \cite{Olivares-DelCampo:2017feq,Boehm:2017dze,Blennow:2019fhy,2106.04446,1804.05117,1803.01773,2110.04024,2112.05057,1909.10478,2108.06928,2110.00021,1608.01307,1601.05798,Penacchioni:2020xhg,Coito:2022kif,Berryman:2022hds,Dev:2022bae,1401.7597,1705.06740,1705.09455,2107.10865,Huang:2021kam,Huang:2018cwo,Farzan:2018pnk,Farzan:2019yvo,2205.12950,1908.02278,2203.11642,Mangano:2006mp} from several perspectives. One clear disadvantage in most of those is the fact that such interactions are implemented at the Lagrangian level, which thus assumes a fixed spin for the DM and determined interactions with the SM particles. This obscures the general features of $\nu$DM interactions and are very model dependent. A more compelling approach relies on on-shell methods \cite{Elvang:2013cua,Dixon:2013uaa,1709.04891}, a set of techniques that allows us to write down amplitudes for particles of any mass and spin without having to rely on Lagrangians. Moreover, it is already known that amplitudes written in this way contains all possible terms at all orders in low-energy effective theory expansion \cite{1909.10551,2008.09652,Ma:2019gtx,Gu:2020thj,Shadmi:2018xan,Christensen:2018zcq,Durieux:2019siw,Dong:2022mcv,Li:2022tec,Balkin:2021dko}.

In this paper we are interested in applying on-shell techniques to study $\nu$DM interactions, that was not yet pursued in the present literature. More specifically, we will focus on characterising how this interaction might alter the evolution of astrophysical neutrinos, in which the DM background induces a new contribution to the neutrino potential and hence modifies its dispersion relation.

\vspace{1em}

\textit{$\nu$DM Potential} - In order to compute the potential generated by the $\nu$DM interactions, we need first to compute the 4-point scattering amplitude for 2 DM particles and 2 neutrinos. Any massive $n$-point on-shell amplitude may be written in terms of spinor variables $\ket{\bm{p}}$ and $\sket{\bm{p}}$, where $p$ is a 4-momentum and the bold notation indicates that we are dealing with massive particles \cite{1709.04891}. The amplitude is then built as the sum of all possible Lorentz-invariant combinations of spinors, where each term in the amplitude must have exactly $2s$ spinors for each particle with spin $s$. In our case the amplitude is written as $\mathcal{A}\left[\nu_1\bar{\nu}_2\chi_3\bar{\chi}_4\right]$, with $\chi$ ($\bar{\chi}$) the DM (anti) particle and the subscripts being labels for the the 4-momenta. Therefore, it will be a function of $2s_\chi$ spinors $\rket{\bm{3}}$, $2s_\chi$ spinors $\rket{\bm{4}}$, one $\rket{\bm{1}}$ and one $\rket{\bm{2}}$, where $s_\chi$ is the spin of the DM and we use the notation $\rket{\bm{p}}$ to denote either $\ket{\bm{p}}$ or $\sket{\bm{p}}$. For example, when $s_\chi=0$ the amplitude is explicitly given by \cite{1709.04891,2008.09652}
\al{\label{eq:schi=0}
\mathcal{A}\left[\nu_1\bar{\nu}_2\chi_3\bar{\chi}_4\right]_{s_\chi=0}  = ~& g_1 \braket{\bm{12}} + g_2 \sbraket{\bm{12}} +\\
& + g_3 \asbraket{\bm{1}}{\bm{3}}{\bm{2}}+ g_4 \sabraket{\bm{1}}{\bm{3}}{\bm{2}},
}
where $g_n$ are independent coefficients that act as couplings and can also be functions of the relevant Mandelstam variables $s_{ij}=(p_i+p_j)^2$. In order to write down Eq. \eqref{eq:schi=0}, we are assuming that there is only one species of DM and neutrinos. In what follows we assume for simplicity just one type of DM particle, while neutrino flavor can be implemented as in Ref. \cite{2103.16362}, such that the spinor variables and the coefficients acquire a dependence on the particular mass-eigenstate considered:
\al{
\mathcal{A}\left[\nu_1\bar{\nu}_2\chi_3\bar{\chi}_4\right] &\rightarrow \mathcal{A}[\nu_1^i \bar{\nu}_2^j\chi_3\bar{\chi}_4]\\
\rket{\bm{1}}\rightarrow \rket{\bm{1}_i},&\quad \rket{\bm{2}}\rightarrow \rket{\bm{2}_j}\\ 
g_n & \rightarrow g_n^{ij}.
} 
For more details on on-shell methods and the conventions used, we refer the reader to the appendix.

For $s_\chi=0$, we need only 4 Lorentz-invariant structures to compute the 4-point amplitude in Eq. \eqref{eq:schi=0}, whereas according to Ref. \cite{2008.09652} the number of independent spinor structures is in general $4\times(2s_\chi+1)^2$ and thus grows quickly with the spin of the DM. Although it becomes inefficient to determine explicitly the amplitudes for general $s_\chi$, we can look out for some patterns that will be useful later. We identify 5 different structures. The first one is
\be\label{eq:I}
\rbraket{\bm{1}_i\bm{2}_j}\rbraket{\bm{34}}^{2s_\chi},
\ee
where $\rbraket{\bm{pk}}=\sbraket{\bm{pk}}$ or $\braket{\bm{pk}}$, and $\rbraket{\bm{34}}^{2s_\chi}$ means that there are $2s_\chi$ independent powers of $\rbraket{\bm{34}}$. The second category is similar to the one before, but with an extra insertion of the DM momentum:
\be\label{eq:II}
\rbraket{\bm{34}}^{2s_\chi}\rrbraket{\bm{1}_i}{\bm{3}}{\bm{2}_j},
\ee
in which we have defined $\rrbraket{\bm{p}}{\bm{k}}{\bm{q}}$ as $\asbraket{\bm{p}}{\bm{k}}{\bm{q}}$ or $\sabraket{\bm{p}}{\bm{k}}{\bm{q}}$. The third type of structure is represented by
\be\label{eq:III}
\rbraket{\bm{34}}^{2s_\chi-1}\braket{\bm{1}_i\bm{3}}\sbraket{\bm{2}_j\bm{4}}~~\text{or}~~\rbraket{\bm{34}}^{2s_\chi-1}\braket{\bm{1}_i\bm{3}}\braket{\bm{2}_j\bm{4}},
\ee
together with their conjugate structures. One remark about Eq. \eqref{eq:III} is that, since massive amplitudes must be regular functions of the spinor variables, the exponent of $\rbraket{\bm{34}}^{2s_\chi-1}$ must not be negative, that is $s_\chi \geq 1/2$. This means that this type of structure can only appear in amplitudes involving DM particles with spin higher than 1/2. The last two categories are similar to the ones of Eqs. \eqref{eq:I} and \eqref{eq:III}, but with a insertion of neutrino momentum:
\be\label{eq:IV}
\rbraket{\bm{1}_i\bm{2}_j}\rrbraket{\bm{3}}{\bm{1}_i}{\bm{4}}\rbraket{\bm{34}}^{2s_\chi-1},
\ee
\vspace{0.05cm}
\al{\label{eq:V}
&\rbraket{\bm{34}}^{2s_\chi-2}\rrbraket{\bm{3}}{\bm{1}_i}{\bm{4}}\braket{\bm{1}_i\bm{3}}\sbraket{\bm{2}_j\bm{4}}~~\text{or}\\
&\rbraket{\bm{34}}^{2s_\chi-2}\rrbraket{\bm{3}}{\bm{1}_i}{\bm{4}}\braket{\bm{1}_i\bm{3}}\braket{\bm{2}_j\bm{4}}.
}
As before, the terms in Eqs. \eqref{eq:IV} and \eqref{eq:V} are only allowed for DM spins higher than $s_\chi\geq 1/2$ and $s_\chi\geq 1$, respectively. It is also important to notice that we only consider structures with at most 1 momentum insertion, as terms with more insertions are redundant in a 4-point amplitude\footnote{The only exception would be terms like $\langle\bm{3}|\bm{12}|\bm{3}\rangle$ that appear for DM spins larger than 1. However, in the case we consider they will turn out to be of higher order in the neutrino mass expansion, so we neglect them.} \cite{2008.09652}.

The appropriate combination of structures \eqref{eq:I}-\eqref{eq:V} yields the most general 4-point amplitude $\mathcal{A}[\nu_1^i \bar{\nu}_2^j\chi_3\bar{\chi}_4]$ for a DM particle of spin $s_\chi$. From this amplitude, we can now follow the steps described in Ref. \cite{2103.16362} to compute the effective potential generated by a DM background. This can be done in three steps: ($i$) take the elastic limit $p_1 \simeq -p_2$ and $p_3=-p_4$ while keeping in general $i\neq j$; ($ii$) average over the DM spins; ($iii$) integrate over the DM distribution profile. More precisely, based on this procedure we can write
\begin{widetext}
\al{\label{eq:potential_procedure}
\mathcal{A}[\nu_1^i \bar{\nu}_2^j\chi_3\bar{\chi}_4]^{I,J, \{K_1,\cdots, K_{2s_\chi}\},\{L_1,\cdots, L_{2s_\chi}\}} & \rightarrow \\
 \rightarrow\frac{\epsilon_{K_1L_1}\cdots\epsilon_{K_{2s_\chi}L_{2s_\chi}}}{(2s_\chi+1)}&\int \frac{\dd^3p_3}{(2\pi)^32E(\vec{p}_3)}\ f_\chi(\vec{p}_3)\mathcal{A}[\nu_1^i \bar{\nu}_{-1}^j\chi_3\bar{\chi}_{-3}]^{I,J, \{K_1,\cdots, K_{2s_\chi}\},\{L_1,\cdots, L_{2s_\chi}\}} ,
}
\end{widetext}
where we are writing explicitly the $SU(2)$ little-group indices $\{I,J,K,L\}$, $f_\chi$ is the DM distribution function in momentum space and $E(\vec{p}_3)$ is the DM energy. Since we are averaging over the DM little-group indices, Eq. \eqref{eq:potential_procedure} can be written in general as
\al{\label{eq:potential_reduction}
\mathcal{A}\left[\nu_1\bar{\nu}_2\chi_3\bar{\chi}_4\right]\to & V_{m}\braket{\bm{1}_i(\bm{-1})_j}+V_{m}'\sbraket{\bm{1}_i(\bm{-1})_j}+\\
& + \asbraket{\bm{1}_i}{V_p}{(\bm{-1})_j}+\sabraket{\bm{1}_i}{V_p'}{(\bm{-1})_j},
}
such that, according to Ref. \cite{2103.16362}, $V_{m}$, $V_{m}'$, $V_p$ and $V_p'$ can be identified as the DM-induced potentials. Notice that we assume that only Standard Model neutrinos are propagating, which thus implies that the little-group indices of the neutrino spinors are essentially fixed to be $\ket{1^I_i}\simeq \ket{1^1_i}$ and $\sket{(-1)^J_j}\simeq \sket{(-1)^2_j}$. In addition to this, we only consider first order terms in the potentials and in the mass of the neutrinos. As a consequence, the term $\sabraket{\bm{1}_i}{V_p'}{(\bm{-1})_j}$ in Eq. \eqref{eq:potential_reduction} is of higher order in the neutrino masses and will from now on be neglected. For more details we refer the reader to Ref. \cite{2103.16362}.

Let us now understand how each of the structures in Eqs. \eqref{eq:I}-\eqref{eq:V} contribute to the potentials $V_m$, $V_m'$ and $V_p$. Performing the spin average over the little-group indices and taking the elastic limit we obtain
\al{
\rbraket{\bm{1}_i\bm{2}_j}\rbraket{\bm{34}}^{2s_\chi} & \rightarrow m_\chi^{2s_\chi}\rbraket{\bm{1}_i(-\bm{1})_j}\\
\rbraket{\bm{34}}^{2s_\chi}\rrbraket{\bm{1}_i}{\bm{3}}{\bm{2}_j} & \rightarrow m_\chi^{2s_\chi}\asbraket{\bm{1}_i}{\bm{3}}{(\bm{-1})_j}\\
\rbraket{\bm{34}}^{2s_\chi-1}\braket{\bm{1}_i\bm{3}}\sbraket{\bm{2}_j\bm{4}} & \rightarrow m_\chi^{2s_\chi-1}\asbraket{\bm{1}_i}{\bm{3}}{(\bm{-1})_j}\\
\rbraket{\bm{34}}^{2s_\chi-1}\braket{\bm{1}_i\bm{3}}\braket{\bm{2}_j\bm{4}} & \rightarrow m_\chi^{2s_\chi}\rbraket{\bm{1}_i(-\bm{1})_j}\\
\rbraket{\bm{1}_i\bm{2}_j}\rrbraket{\bm{3}}{\bm{1}_i}{\bm{4}}\rbraket{\bm{34}}^{2s_\chi-1} &\rightarrow m_\chi^{2s_\chi-1}(p_3\cdot p_1)\rbraket{\bm{1}_i(-\bm{1})_j}\\
\rbraket{\bm{34}}^{2s_\chi-2}\rrbraket{\bm{3}}{\bm{1}_i}{\bm{4}}\braket{\bm{1}_i\bm{3}}&\sbraket{\bm{2}_j\bm{4}} \rightarrow\\
\rightarrow &~ m_\chi^{2s_\chi-2}(p_3\cdot p_1)\asbraket{\bm{1}_i}{\bm{3}}{(\bm{-1})_j}\\
\rbraket{\bm{34}}^{2s_\chi-2}\rrbraket{\bm{3}}{\bm{1}_i}{\bm{4}}\braket{\bm{1}_i\bm{3}}&\braket{\bm{2}_j\bm{4}}\rightarrow\\
\rightarrow &~ m_\chi^{2s_\chi-1}(p_3\cdot p_1)\rbraket{\bm{1}_i(-\bm{1})_j}.
}
The integration over the DM distribution function depends not only on the particular form of $f_\chi$ but also on how the coefficients of the amplitude depend on $p_3$. As a first and reasonable approximation, then, we take into consideration the non-relativistic nature of the DM and assume $f_\chi(\vec{p}_3) = N_\chi \delta^{(3)}(\vec{p}_3)$, with $N_\chi$ the number density of DM particles. Under this assumption, $p_3/E(\vec{p}_3)\rightarrow (1,\vec{0})$ and $p_3\cdot p_1 \rightarrow m_\chi E_\nu$, where $E_\nu$ is the neutrino energy. The potentials in Eq. \eqref{eq:potential_reduction} are therefore given by
\al{\label{eq:partial}
V_m^{(')} &= N_\chi m_\chi^{2s_\chi-1}\left(c_1^{(')} + c_2^{(')} E_\nu\right)\\
V_p &= N_\chi m_\chi^{2s_\chi-1}\left(c_3 + c_4 m_\chi + c_5 E_\nu\right).
}
The coefficients $c_{n}$ are combinations of the couplings of the initial 4-point amplitude and may depend on $E_\nu$, $m_\chi$ and $m_\nu$ as well. We emphasise again that for $s_\chi=0$ we have $c_2^{(')}=c_3=c_5=0$, while for $s_\chi=1/2$ we have $c_5=0$. It is also interesting to notice that to derive the same results using only quantum field theoretical techniques would be an arduous task, since one would need to explicitly list all possible ways of generating $\mathcal{A}[\nu_1^i \bar{\nu}_2^j\chi_3\bar{\chi}_4]$ for each DM spin. For higher spins this would be even more challenging, as in some cases it is impossible to have a Lagrangian description. By means of on-shell methods, we can completely bypass such hurdles and straightforwardly obtain Eq. \eqref{eq:partial}.

We see from expression \eqref{eq:partial} that $c_n$ must be dimensionful in order to produce potentials with the correct dimension. More precisely, the coefficients associated to structures with no momentum insertion ($c_1$ and $c_4$) have mass dimension $(2s_\chi+1)^{-1}$, whereas the ones related to terms with one momentum insertion ($c_2$, $c_3$ and $c_5$) have dimension $(2s_\chi+2)^{-1}$. The physical interpretation of this mass dimension is nevertheless not unique. On the one hand, if we take the amplitude $\mathcal{A}[\nu_1^i \bar{\nu}_2^j\chi_3\bar{\chi}_4]$ to be a contact interaction, \textit{i.e.} with the couplings being regular functions of the Mandelstam variables, then we can introduce a large scale $\Lambda^2\gg |s_{13}|,m_\chi^2$ to correct the dimensionality of the couplings:
\al{\label{eq:EFT}
V_m^{(')} &= \frac{N_\chi}{\Lambda^2} \left(\frac{m_\chi}{\Lambda}\right)^{2s_\chi-1}\left[\hat{c}_1^{(')} + \hat{c}_2^{(')} \frac{E_\nu}{\Lambda}\right]\\
V_p &= \frac{N_\chi}{\Lambda^2} \left(\frac{m_\chi}{\Lambda}\right)^{2s_\chi-1}\left[\hat{c}_3 + \hat{c}_4 \frac{m_\chi}{\Lambda} + \hat{c}_5 \frac{E_\nu}{\Lambda}\right],
}
where now each (dimensionless) $\hat{c}_n$ is to be interpreted as an infinite expansion in powers of $s_{13}/\Lambda^2$ and $m_\chi/\Lambda$. This resembles very much the usual Effective Field Theory (EFT) approach, in which we have a cut-off scale that sets the maximum energy scale allowed and we can expand observables in terms of inverse powers of the cut-off. Though similar, we cannot directly identify $\Lambda$ as this cut-off scale, because at the level of amplitudes there is no way to know what is the scaling of the coefficients with the cut-off scale. It is also important to notice that the dimensionality of $c_n$ in Eq. \eqref{eq:partial} grows with $s_\chi$. This is naturally related to the fact that higher spin particles can only interact via effective interactions and will thus be suppressed by extra powers of some other cut-off scale \cite{1712.02346,1811.01952,2010.02224,2011.10058}. In Eq. \eqref{eq:EFT}, $\Lambda$ takes into account both the EFT and the higher spin cut-off scales in some non-trivial way \footnote{Here we always assume that the expasion of the coefficients in powers of $\Lambda$ begins with a $\Lambda$-independent term.}. On the other hand, the coefficients may also posses poles and branch-cuts, which correspond respectively to tree and loop diagrams. For instance, if we consider a light particle $\phi$ being exchanged at tree-level, then the coefficients will have a contribution given by $(s_{ij}-m_\phi^2)^{-1}$, where in the elastic limit $s_{12}\simeq 0$ and $s_{13},s_{14}\simeq m_\chi^2\pm 2m_\chi E_\nu$. As a consequence, instead of a suppression by $\Lambda^2$, we can have an enhancement to the potential coming from the small masses $m_\chi^2$ and $m_\phi^2$. This effect, however, strongly depend on the mass of $\phi$ and on its spin, as for a given choice of DM spin not all channels are allowed. As for loop effects, we also expect some degree of enhancement compared to Eq. \eqref{eq:EFT} but will be much more model dependent than the tree-level case, so we do not pursue them further in this work.

\begin{figure}[tb]
\centering
\includegraphics[width=0.47\textwidth]{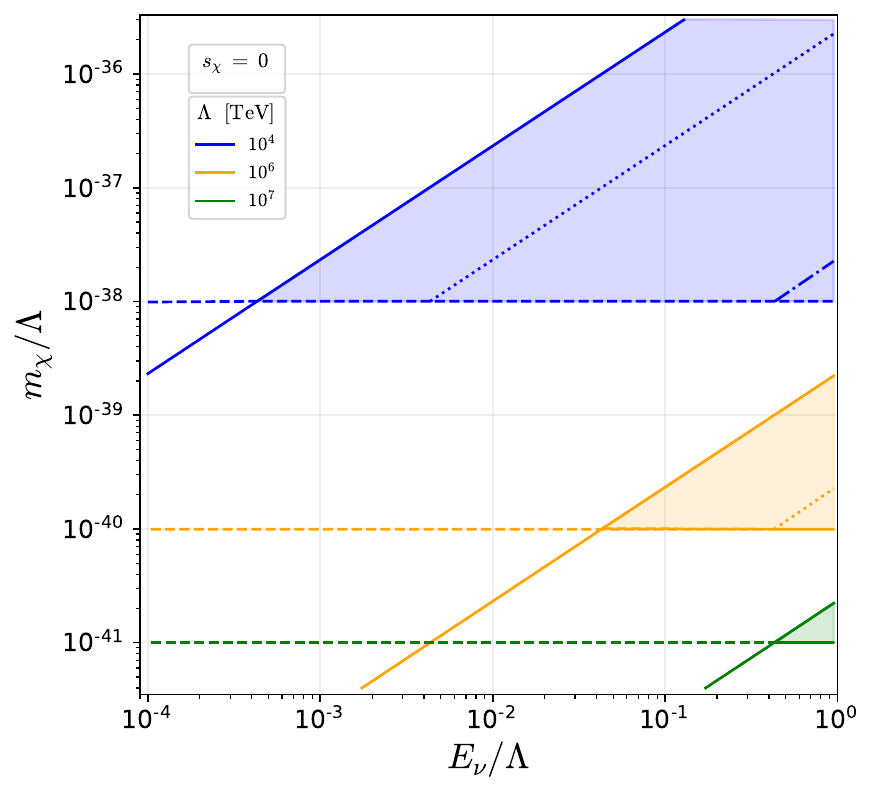}
\vglue -0.3cm
\caption{Regions in the $E_\nu/\Lambda\times m_\chi/\Lambda$ plane in which $V_p/V_0\gtrsim 1$ (below the solid lines) and $m_\chi \gtrsim 10^{-22}$ eV (above the dashed lines) are satisfied simultaneously for $\Lambda= 10^4$ TeV (blue), $10^6$ TeV (orange) and $10^7$ TeV (green). The $\nu$DM potential is taken to be the contact one in Eq. \eqref{eq:EFT} and we choose $s_\chi=0$. Below the dotted (dashed-dotted) lines we have $V_p/V_0\gtrsim 10$ ($10^3$).}\label{fig:results}
\vglue -0.3cm
\end{figure}

\vspace{1em}

\textit{Phenomenology} - Having determined the general form for the DM potentials in Eq. \eqref{eq:partial}, we now investigate under what circumstances they could give an observable signal. We first notice that the potentials $V_m^{(')}$ and $V_p$ contribute in distinct ways to the total neutrino potential: $V_m^{(')}$ is a correction to the neutrino masses $m_\nu$, while $V_p$ corrects the vacuum-energy $V_0\equiv \Delta m_\nu^2/2E_\nu$, with $\Delta m_\nu^2$ refering to the usual mass-squared splittings $\Delta m^2_{\nu ij}\equiv m_i^2-m_j^2$ \cite{Giunti:1053706,barger2012physics,ParticleDataGroup:2020ssz}. This can be seen, as shown in Ref. \cite{2103.16362}, by resumming the neutrino propagator in the presence of this background and computing the new dispersion relation of the neutrino:
\be
p^2-m_\nu^2\to (p-V_p)^2 - m_\nu^2 - m_\nu V_m - V_m'm_\nu,
\ee
where $m_\nu$ is to be understood as the neutrino mass matrix. Since usually experiments are much more sensitive to the vacuum-energy rather than the absolute mass scales, we will consider only $V_p$ to be of relevance, but the discussion would be analogous for $V_m^{(')}$ \footnote{Given that in general $\Delta m_\nu^2/2E_\nu$ is much smaller than the absolute neutrino mass scale for high neutrino energies, $V_m^{(')}$ will have to be much larger than $V_p$ in order to present similar effects.}. The impact of $V_p$ to neutrino observables is similar to that of the standard Mikheyev-Smirnov-Wolfenstein potential \cite{Wolfenstein:1977ue,Mikheyev:1985zog,Barger:1980tf}, namely to affect the flavor composition of the incoming neutrino flux, in which the mixing angles and mass splittings are modified as functions $V_p/V_0$. With this ratio we can quantify how large the DM potential is compared to the oscillations in vacuum, that are dictated by $V_0$. The precise way in which the observables are altered depend naturally on the flavor structure of the potential, which in the present framework is completely arbitrary. The dependence on the strength of the potential, however, will always be parameterised by $V_p/V_0$ since we need to recover the usual oscillation theory as $V_p/V_0\to 0$. Hence, by estimating this ratio we can already have a good measure of how much the flavor structure is expected to change in the presence of the DM background. In the following we analyse the case of astrophysical neutrinos, that have energies of order PeV, propagating in the DM halo of the Milky-Way, as these neutrinos are the ones that experience at most this background.

Let us first consider the potential generated by contact interactions given in Eq. \eqref{eq:EFT}. In this case we must impose that $\Lambda\gg E_\nu,m_\chi$ in order to have well defined expansion parameters. We thus obtain for the leading term of $V_p$
\begin{widetext}
\be\label{eq:EFT_estimate}
\frac{V_p}{V_0}\simeq \left(\frac{\rho_\chi}{\rho_0}\right)\left(\frac{E_\nu}{\Lambda}\right)^2\left(\frac{\text{PeV}}{E_\nu}\right)\left(\frac{\Delta m_{12}^2}{\Delta m_\nu^2}\right)\left(\frac{10^{-18}~\text{eV}}{m_\chi}\right)\left\{\begin{array}{ll}
\left(\frac{m_\chi}{\Lambda}\right)^{2s_\chi-1},~~s_\chi\geq 1\\
~~~1,\qquad\qquad s_\chi \leq 1/2 ,\\
\end{array}\right.
\ee
\end{widetext}
up to a order one coefficient that carries the relevant flavor structure. In the above we have considered the largest neutrino mass-squared splitting $\Delta m_{12}^2=2.453\times 10^{-3}~\text{eV}^2$ \cite{ParticleDataGroup:2020ssz}. Choosing $\Delta m_{13}^2$ instead would have changed the estimate above by a factor of $10^{2}$, but would not qualitatively change our conclusions. The precise combination of mass splittings appearing in Eq. \eqref{eq:EFT_estimate} depend on how the DM couples to neutrino flavors. Also, $\rho_\chi=m_\chi N_\chi$ is the DM energy density and $\rho_0=0.39~\text{GeV}/\text{cm}^3$ is the local energy density, while the term in brackets depend on the spin of the DM, as we remember that some spinor structures are absent for $s_\chi=0$ and 1/2. From Eq. \eqref{eq:EFT_estimate} we can now draw some conclusions. Take for instance the case $s_\chi=0$ with $\rho_\chi=\rho_0$\footnote{Varying $\rho_\chi$ according to standard DM profiles for the Milky-Way does not lead to significant changes to the final results.} and $\hat{c}_4=1$. We see that in order to obtain ratios at least of order 1, we need extremely light DM, that essentially lie in the fuzzy regime \cite{astro-ph/0003365}. This agrees with what is usually found in the literature \cite{Farzan:2018pnk,1909.10478}. In Fig. \ref{fig:results} we show the regions that satisfy at the same time $V_p/V_0\gtrsim 1$ and $m_\chi \gtrsim 10^{-22}$ eV, which is the lower bound for an ultra light DM \cite{astro-ph/0003365,Ferreira:2020fam}, for several values of the scale $\Lambda$. The values of $\Lambda$ are chosen such to satisfy the requirement $\Lambda>E_\nu\sim$ PeV. The region of interest is then concentrated at DM mass values $m_\chi\lesssim 10^{-19}$ eV and for larger values of $E_\nu/\Lambda$. From Eq. \eqref{eq:EFT_estimate} it is clear that in the case $s_\chi=1/2$ the leading term is insensitive to the small ratio $m_\chi/\Lambda$ as well, so we could in principle expect a similar signal in this case. Notwithstanding, the mass ranges allowed violate by far the Tremaine-Gunn bound, which forbids fermionic DM with masses less than $\sim$1 eV. For higher spins we must take the same result and multiply by the appropriate power of $m_\chi/\Lambda$, that as we can see would completely suppress $V_p/V_0$ and make this effect unobservable. We thus arrive at the conclusion that, in the case of contact $\nu$DM interactions, no DM other than the scalar one can produce a phenomenologically relevant potential for propagating astrophysical neutrinos. Moreover, the masses of the scalar DM must all be in the range $10^{-22}~\text{eV}\lesssim m_\chi \lesssim 10^{-19}$ eV to produce an order 1 ratio $V_p/V_0$.

It is worth remarking that although in the case of contact $\nu$DM interactions it is not possible to have a relevant signal with fermionic mediators, we can still rely on the enhancement at tree-level to obtain a more favourable picture. In order to satisfy both $V_p/V_0\gtrsim 1$ and $m_\chi \gtrsim 1$ eV at the same time, we would need a mediator mass at least as low as 1 MeV. In this case, however, due to the mediator being very light, it is not trivial to evade bounds from cosmology as the rate of the interactions might be efficient in the Early Universe. To precisely derive such bounds we would need to specify how exactly the neutrinos interact with the DM, which is not the purpose of the present paper. In the case of contact amplitudes and for the relevant scales we are interested in, $\Lambda\gtrsim 10^4$ TeV, the $\nu$DM interaction is extremely suppressed and we thus do not expect any relevant modification to the cosmological history.

The discussion above can be generalized in several directions. As a first step, one could consider adding sterile neutrinos that can also interact with the DM. Such a scenario is a straightforward generalization of the present discussion with, however, a more involed flavor structure and model dependency. In general we expect, as before, changes in the flavor composition of the incoming neutrino flux. Another extension of the present work is to consider the effects of the $\nu$DM scattering, which includes changes in the energy and angular spectra \cite{Arguelles:2015dca,Arguelles:2017atb}, the cosmological history \cite{Mangano:2006mp,1205.5809} and possibly oscilation damping \cite{Stodolsky:1986dx}. A particular case in which the analysis of the $\nu$DM scattering could be useful is the one of active splitting \cite{Davoudiasl:2018hjw}, \textit{i.e.} when the neutrino masses are dynamically generated by the $\nu$DM potential $V_p$ itself. This is so, because under the assumption that the $\nu$DM potential exactly mimics the standard flavor structure, we might need other observables to probe the $\nu$DM interaction. We leave the discussion of these points to future work.

\vspace{1em}

\textit{Conclusions} - In this paper we have studied how an interaction between DM and neutrinos can affect the propagation of the latter in a DM background. Employing on-shell methods, we have managed to obtain a general characterisation of DM-induced potentials in a completely model independent way. Our findings can be summarised in three points: ($i$) Eq. \eqref{eq:partial} represents, to first order in couplings and neutrino masses, the most general potentials that a DM background can produce; ($ii$) for contact $\nu$DM interactions, we obtain in a novel and more general way the result that only ultra light scalar DM could give an observable signal at the level of 4-point amplitude; ($iii$) DM spins larger than 1/2 cannot produce a relevant ratio $V_p/V_0$ unless some degree of enchament is present, for example due to light particles being exchanged. Future measurements of the astrophysical flavor composition by IceCube will therefore be able to probe not only the strength and the flavor structure of $\nu$DM interactions, but the nature of DM itself as well.

\begin{acknowledgements}
The author thanks Enrico Bertuzzo, Joachim Kopp, Pedro Bittar and Gustavo F. S. Alves for useful discussions, and acknowledges financial support from Fundação de Amparo à Pesquisa de São Paulo (FAPESP) under contract 2020/14713-2.
\end{acknowledgements}

\vspace{1em}

\textit{Appendix} - In this appendix we review some of the key concepts of on-shell methods and set the notation for spinor variables. For a given 4-momentum $p$, the spinor variables are given by
\be\label{eq:massive_helicity_up}
\bra{p^{I}}^\alpha = -
\left(
\begin{array}{c|c}
\sqrt{E+|\vec{p}|} \,c & - \sqrt{E-|\vec{p}|}\, s \\
\sqrt{E+|\vec{p}|}\, s^* & \sqrt{E-|\vec{p}|} \,c
\end{array}\right),
\ee
\be
\sket{p^{ I}}^{\dot{\alpha}} = 
\left(\begin{array}{c|c}
(\sqrt{E-|\vec{p}|}\, s)^* & (\sqrt{E+|\vec{p}|} \,c)^* \\
- (\sqrt{E-|\vec{p}|}\, c)^* & (\sqrt{E+|\vec{p}|}\, s^*)^*
\end{array}\right) ,
\ee
where $I=1,2$ is the $SU(2)$ litle-group index of the Lorentz group (columns) and $\alpha,\dot{\alpha}$ are the indices of the $(1/2,0)$ and $(0,1/2)$ representations (lines), respectively. In addition, $c=\cos(\theta/2)$ and $s=\sin(\theta/2)e^{i\phi}$ are the angular variables, $\vec{p}=|\vec{p}|(\sin{\theta}\cos{\phi},\sin{\theta}\sin{\phi},\cos{\theta})$ is the 3-momentum and $E$ the energy. In general, we allow for the 4-momenta to be complex. We can also define the down-index variables using the Levi-Civita tensor $\epsilon$, for instance as $\sbra{p^I}_{\dot{\alpha}} = \sket{p^{ I}}^{\dot{\beta}}\epsilon_{\dot{\beta}\dot{\alpha}}$, with $\epsilon^{12}=-\epsilon_{12}=1$, and similarly for the angle variable. All the spinors transform in the fundamental of the $SU(2)$ little-group:
\be
\rket{p^I}\rightarrow W^I_{~J}\rket{p^J},\quad W\in SU(2),
\ee
where $\rket{p^I}$ is used to denote both types of spinors and, since the $S$-matrix elements are little-group covariant, they can be therefore used as building blocks for constructing amplitudes. Scattering amplitudes are thus written as the sum of all possible Lorentz invariant spinor structures with the correct little-group transformation. More precisely, amplitudes involving a particle of spin $s$ can be represented as a $2s$ symmetric tensor of the corresponding little-group. We employ the bold notatin, in which we suppress the little-group index and bold the momentum inside the spinor, \textit{e.g.} $\ket{p^I}\to \ket{\bm{p}}$, while leaving implicit symmetrization over all little-group indices \cite{1709.04891}.

We can define the Lorentz invariant angle and square products between two spinors as
\be
\braket{p^Iq^J} \equiv \bra{p^I}^\alpha \ket{q^J}_\alpha,\quad \sbraket{p^Iq^J} \equiv \sbra{p^I}_{\dot{\alpha}} \sket{q^J}^{\dot{\alpha}}.
\ee
In this notation, the 4-momentum $p$ can be written as a bispinor:
\be
p^{\dot{\alpha}\alpha} = \epsilon_{IJ} \sket{p^I}^{\dot{\alpha}}\bra{p^J}^\alpha,\quad p_{\alpha\dot{\alpha}} = -\epsilon_{IJ} \ket{p^I}_{{\alpha}}\sbra{p^J}_{\dot{\alpha}}.
\ee
The spinor variables all satisfy the Weyl equations
\al{\label{eq:Weyl}
p\sket{p^I} = m \ket{p^I},&\quad p\ket{p^I} = m \sket{p^I},\\
\sbra{p^I}p = -m \bra{p^I},&\quad \bra{p^I}p =- m \sbra{p^I},
}
with $m$ the mass of the particle. From the above equations we can deduce that
\al{
\braket{p^Ip^J}=m~\epsilon^{IJ},&\quad \sbraket{p^Ip^J}=-m~\epsilon^{IJ},\\
\ket{p^I}_\alpha \bra{p_I}^\beta = m~\delta_\alpha^\beta,&\quad \sket{p^I}^{\dot{\alpha}} \sbra{p_I}_{\dot{\beta}} = -m~\delta^{\dot{\alpha}}_{\dot{\beta}}.
}
Finally, we can define spinor variables with negative momentum. The correct analytic continuation that still respect the Weyl equations \eqref{eq:Weyl} is
\be
\ket{(-p)^I}=\ket{p^I}, \quad \sket{(-p)^I}=-\sket{p^I}.
\ee

\bibliographystyle{JHEP2}
\bibliography{nuDM_Potential_Paper}% Produces the bibliography via BibTeX.

\providecommand{\href}[2]{#2}\begingroup\raggedright\begin{thebibliography}{10}

\bibitem{1304.5356}
{\scshape IceCube} collaboration, \emph{{First observation of PeV-energy
  neutrinos with IceCube}},
  \href{https://doi.org/10.1103/PhysRevLett.111.021103}{\emph{Phys. Rev. Lett.}
  {\bfseries 111} (2013) 021103}
  [\href{https://arxiv.org/abs/1304.5356}{{\ttfamily 1304.5356}}].

\bibitem{Olivares-DelCampo:2017feq}
A.~Olivares-Del~Campo, C.~B\oe{}hm, S.~Palomares-Ruiz and S.~Pascoli,
  \emph{{Dark matter-neutrino interactions through the lens of their
  cosmological implications}},
  \href{https://doi.org/10.1103/PhysRevD.97.075039}{\emph{Phys. Rev. D}
  {\bfseries 97} (2018) 075039}
  [\href{https://arxiv.org/abs/1711.05283}{{\ttfamily 1711.05283}}].

\bibitem{Boehm:2017dze}
C.~Boehm, A.~Olivares-Del~Campo, S.~Palomares-Ruiz and S.~Pascoli,
  \emph{{Phenomenology of a Neutrino-DM Coupling: The Scalar Case}},  in
  \emph{{Prospects in Neutrino Physics}}, 5, 2017,
  \href{https://arxiv.org/abs/1705.03692}{{\ttfamily 1705.03692}}.

\bibitem{Blennow:2019fhy}
M.~Blennow, E.~Fernandez-Martinez, A.~Olivares-Del~Campo, S.~Pascoli,
  S.~Rosauro-Alcaraz and A.~V. Titov, \emph{{Neutrino Portals to Dark Matter}},
  \href{https://doi.org/10.1140/epjc/s10052-019-7060-5}{\emph{Eur. Phys. J. C}
  {\bfseries 79} (2019) 555}
  [\href{https://arxiv.org/abs/1903.00006}{{\ttfamily 1903.00006}}].

\bibitem{2106.04446}
C.~Gu\'epin, R.~Aloisio, L.~A. Anchordoqui, A.~Cummings, J.~F. Krizmanic, A.~V.
  Olinto et~al., \emph{{Indirect dark matter searches at ultrahigh energy
  neutrino detectors}},
  \href{https://doi.org/10.1103/PhysRevD.104.083002}{\emph{Phys. Rev. D}
  {\bfseries 104} (2021) 083002}
  [\href{https://arxiv.org/abs/2106.04446}{{\ttfamily 2106.04446}}].

\bibitem{1804.05117}
F.~Capozzi, I.~M. Shoemaker and L.~Vecchi, \emph{{Neutrino Oscillations in Dark
  Backgrounds}},
  \href{https://doi.org/10.1088/1475-7516/2018/07/004}{\emph{JCAP} {\bfseries
  07} (2018) 004} [\href{https://arxiv.org/abs/1804.05117}{{\ttfamily
  1804.05117}}].

\bibitem{1803.01773}
J.~Liao, D.~Marfatia and K.~Whisnant, \emph{{Light scalar dark matter at
  neutrino oscillation experiments}},
  \href{https://doi.org/10.1007/JHEP04(2018)136}{\emph{JHEP} {\bfseries 04}
  (2018) 136} [\href{https://arxiv.org/abs/1803.01773}{{\ttfamily
  1803.01773}}].

\bibitem{2110.04024}
D.~C. Hooper and M.~Lucca, \emph{{Hints of dark matter-neutrino interactions in
  Lyman-\ensuremath{\alpha} data}},
  \href{https://doi.org/10.1103/PhysRevD.105.103504}{\emph{Phys. Rev. D}
  {\bfseries 105} (2022) 103504}
  [\href{https://arxiv.org/abs/2110.04024}{{\ttfamily 2110.04024}}].

\bibitem{2112.05057}
E.~J. Chun, \emph{{Neutrino Transition in Dark Matter}},
  \href{https://arxiv.org/abs/2112.05057}{{\ttfamily 2112.05057}}.

\bibitem{1909.10478}
K.-Y. Choi, E.~J. Chun and J.~Kim, \emph{{Neutrino Oscillations in Dark
  Matter}}, \href{https://doi.org/10.1016/j.dark.2020.100606}{\emph{Phys. Dark
  Univ.} {\bfseries 30} (2020) 100606}
  [\href{https://arxiv.org/abs/1909.10478}{{\ttfamily 1909.10478}}].

\bibitem{2108.06928}
D.~Green, D.~E. Kaplan and S.~Rajendran, \emph{{Neutrino interactions in the
  late universe}}, \href{https://doi.org/10.1007/JHEP11(2021)162}{\emph{JHEP}
  {\bfseries 11} (2021) 162}
  [\href{https://arxiv.org/abs/2108.06928}{{\ttfamily 2108.06928}}].

\bibitem{2110.00021}
D.~Borah, M.~Dutta, S.~Mahapatra and N.~Sahu, \emph{{Self-interacting dark
  matter via right handed neutrino portal}},
  \href{https://doi.org/10.1103/PhysRevD.105.015004}{\emph{Phys. Rev. D}
  {\bfseries 105} (2022) 015004}
  [\href{https://arxiv.org/abs/2110.00021}{{\ttfamily 2110.00021}}].

\bibitem{1608.01307}
A.~Berlin, \emph{{Neutrino Oscillations as a Probe of Light Scalar Dark
  Matter}}, \href{https://doi.org/10.1103/PhysRevLett.117.231801}{\emph{Phys.
  Rev. Lett.} {\bfseries 117} (2016) 231801}
  [\href{https://arxiv.org/abs/1608.01307}{{\ttfamily 1608.01307}}].

\bibitem{1601.05798}
P.~F. de~Salas, R.~A. Lineros and M.~T\'ortola, \emph{{Neutrino propagation in
  the galactic dark matter halo}},
  \href{https://doi.org/10.1103/PhysRevD.94.123001}{\emph{Phys. Rev. D}
  {\bfseries 94} (2016) 123001}
  [\href{https://arxiv.org/abs/1601.05798}{{\ttfamily 1601.05798}}].

\bibitem{Penacchioni:2020xhg}
A.~V. Penacchioni, O.~Civitarese and C.~R. Arg\"uelles, \emph{{Testing dark
  matter distributions by neutrino\textendash{}dark matter interactions}},
  \href{https://doi.org/10.1140/epjc/s10052-020-7744-x}{\emph{Eur. Phys. J. C}
  {\bfseries 80} (2020) 183}.

\bibitem{Coito:2022kif}
L.~Coito, C.~Faubel, J.~Herrero-Garc\'\i{}a, A.~Santamaria and A.~Titov,
  \emph{{Sterile neutrino portals to Majorana dark matter: effective operators
  and UV completions}},
  \href{https://doi.org/10.1007/JHEP08(2022)085}{\emph{JHEP} {\bfseries 08}
  (2022) 085} [\href{https://arxiv.org/abs/2203.01946}{{\ttfamily
  2203.01946}}].

\bibitem{Berryman:2022hds}
J.~M. Berryman et~al., \emph{{Neutrino Self-Interactions: A White Paper}},  in
  \emph{{2022 Snowmass Summer Study}}, 3, 2022,
  \href{https://arxiv.org/abs/2203.01955}{{\ttfamily 2203.01955}}.

\bibitem{Dev:2022bae}
A.~Dev, G.~Krnjaic, P.~Machado and H.~Ramani, \emph{{Constraining Feeble
  Neutrino Interactions with Ultralight Dark Matter}},
  \href{https://arxiv.org/abs/2205.06821}{{\ttfamily 2205.06821}}.

\bibitem{1401.7597}
R.~J. Wilkinson, C.~Boehm and J.~Lesgourgues, \emph{{Constraining Dark
  Matter-Neutrino Interactions using the CMB and Large-Scale Structure}},
  \href{https://doi.org/10.1088/1475-7516/2014/05/011}{\emph{JCAP} {\bfseries
  05} (2014) 011} [\href{https://arxiv.org/abs/1401.7597}{{\ttfamily
  1401.7597}}].

\bibitem{1705.06740}
G.~Krnjaic, P.~A.~N. Machado and L.~Necib, \emph{{Distorted neutrino
  oscillations from time varying cosmic fields}},
  \href{https://doi.org/10.1103/PhysRevD.97.075017}{\emph{Phys. Rev. D}
  {\bfseries 97} (2018) 075017}
  [\href{https://arxiv.org/abs/1705.06740}{{\ttfamily 1705.06740}}].

\bibitem{1705.09455}
V.~Brdar, J.~Kopp, J.~Liu, P.~Prass and X.-P. Wang, \emph{{Fuzzy dark matter
  and nonstandard neutrino interactions}},
  \href{https://doi.org/10.1103/PhysRevD.97.043001}{\emph{Phys. Rev. D}
  {\bfseries 97} (2018) 043001}
  [\href{https://arxiv.org/abs/1705.09455}{{\ttfamily 1705.09455}}].

\bibitem{2107.10865}
M.~Losada, Y.~Nir, G.~Perez and Y.~Shpilman, \emph{{Probing scalar dark matter
  oscillations with neutrino oscillations}},
  \href{https://doi.org/10.1007/JHEP04(2022)030}{\emph{JHEP} {\bfseries 04}
  (2022) 030} [\href{https://arxiv.org/abs/2107.10865}{{\ttfamily
  2107.10865}}].

\bibitem{Huang:2021kam}
G.-y. Huang and N.~Nath, \emph{{Neutrino meets ultralight dark matter:
  0\ensuremath{\nu}\ensuremath{\beta}\ensuremath{\beta} decay and cosmology}},
  \href{https://doi.org/10.1088/1475-7516/2022/05/034}{\emph{JCAP} {\bfseries
  05} (2022) 034} [\href{https://arxiv.org/abs/2111.08732}{{\ttfamily
  2111.08732}}].

\bibitem{Huang:2018cwo}
G.-Y. Huang and N.~Nath, \emph{{Neutrinophilic Axion-Like Dark Matter}},
  \href{https://doi.org/10.1140/epjc/s10052-018-6391-y}{\emph{Eur. Phys. J. C}
  {\bfseries 78} (2018) 922}
  [\href{https://arxiv.org/abs/1809.01111}{{\ttfamily 1809.01111}}].

\bibitem{Farzan:2018pnk}
Y.~Farzan and S.~Palomares-Ruiz, \emph{{Flavor of cosmic neutrinos preserved by
  ultralight dark matter}},
  \href{https://doi.org/10.1103/PhysRevD.99.051702}{\emph{Phys. Rev. D}
  {\bfseries 99} (2019) 051702}
  [\href{https://arxiv.org/abs/1810.00892}{{\ttfamily 1810.00892}}].

\bibitem{Farzan:2019yvo}
Y.~Farzan, \emph{{Ultra-light scalar saving the 3 + 1 neutrino scheme from the
  cosmological bounds}},
  \href{https://doi.org/10.1016/j.physletb.2019.134911}{\emph{Phys. Lett. B}
  {\bfseries 797} (2019) 134911}
  [\href{https://arxiv.org/abs/1907.04271}{{\ttfamily 1907.04271}}].

\bibitem{2205.12950}
{\scshape IceCube} collaboration, \emph{{Searches for Connections between Dark
  Matter and High-Energy Neutrinos with IceCube}},
  \href{https://arxiv.org/abs/2205.12950}{{\ttfamily 2205.12950}}.

\bibitem{1908.02278}
J.~M. Cline, \emph{{Viable secret neutrino interactions with ultralight dark
  matter}}, \href{https://doi.org/10.1016/j.physletb.2019.135182}{\emph{Phys.
  Lett. B} {\bfseries 802} (2020) 135182}
  [\href{https://arxiv.org/abs/1908.02278}{{\ttfamily 1908.02278}}].

\bibitem{2203.11642}
M.~M. Reynoso, O.~A. Sampayo and A.~M. Carulli, \emph{{Neutrino interactions
  with ultralight axion-like dark matter}},
  \href{https://doi.org/10.1140/epjc/s10052-022-10228-w}{\emph{Eur. Phys. J. C}
  {\bfseries 82} (2022) 274}
  [\href{https://arxiv.org/abs/2203.11642}{{\ttfamily 2203.11642}}].

\bibitem{Mangano:2006mp}
G.~Mangano, A.~Melchiorri, P.~Serra, A.~Cooray and M.~Kamionkowski,
  \emph{{Cosmological bounds on dark matter-neutrino interactions}},
  \href{https://doi.org/10.1103/PhysRevD.74.043517}{\emph{Phys. Rev. D}
  {\bfseries 74} (2006) 043517}
  [\href{https://arxiv.org/abs/astro-ph/0606190}{{\ttfamily
  astro-ph/0606190}}].

\bibitem{Elvang:2013cua}
H.~Elvang and Y.-t. Huang, \emph{{Scattering Amplitudes}},
  \href{https://arxiv.org/abs/1308.1697}{{\ttfamily 1308.1697}}.

\bibitem{Dixon:2013uaa}
L.~J. Dixon, \emph{{A brief introduction to modern amplitude methods}},  in
  \emph{{Theoretical Advanced Study Institute in Elementary Particle Physics}:
  {Particle Physics: The Higgs Boson and Beyond}}, pp.~31--67, 2014,
  \href{https://arxiv.org/abs/1310.5353}{{\ttfamily 1310.5353}},
  \href{https://doi.org/10.5170/CERN-2014-008.31}{DOI}.

\bibitem{1709.04891}
N.~Arkani-Hamed, T.-C. Huang and Y.-t. Huang, \emph{{Scattering amplitudes for
  all masses and spins}},
  \href{https://doi.org/10.1007/JHEP11(2021)070}{\emph{JHEP} {\bfseries 11}
  (2021) 070} [\href{https://arxiv.org/abs/1709.04891}{{\ttfamily
  1709.04891}}].

\bibitem{1909.10551}
G.~Durieux, T.~Kitahara, Y.~Shadmi and Y.~Weiss, \emph{{The electroweak
  effective field theory from on-shell amplitudes}},
  \href{https://doi.org/10.1007/JHEP01(2020)119}{\emph{JHEP} {\bfseries 01}
  (2020) 119} [\href{https://arxiv.org/abs/1909.10551}{{\ttfamily
  1909.10551}}].

\bibitem{2008.09652}
G.~Durieux, T.~Kitahara, C.~S. Machado, Y.~Shadmi and Y.~Weiss,
  \emph{{Constructing massive on-shell contact terms}},
  \href{https://doi.org/10.1007/JHEP12(2020)175}{\emph{JHEP} {\bfseries 12}
  (2020) 175} [\href{https://arxiv.org/abs/2008.09652}{{\ttfamily
  2008.09652}}].

\bibitem{Ma:2019gtx}
T.~Ma, J.~Shu and M.-L. Xiao, \emph{{Standard model effective field theory from
  on-shell amplitudes*}},
  \href{https://doi.org/10.1088/1674-1137/aca200}{\emph{Chin. Phys. C}
  {\bfseries 47} (2023) 023105}
  [\href{https://arxiv.org/abs/1902.06752}{{\ttfamily 1902.06752}}].

\bibitem{Gu:2020thj}
J.~Gu and L.-T. Wang, \emph{{Sum Rules in the Standard Model Effective Field
  Theory from Helicity Amplitudes}},
  \href{https://doi.org/10.1007/JHEP03(2021)149}{\emph{JHEP} {\bfseries 03}
  (2021) 149} [\href{https://arxiv.org/abs/2008.07551}{{\ttfamily
  2008.07551}}].

\bibitem{Shadmi:2018xan}
Y.~Shadmi and Y.~Weiss, \emph{{Effective Field Theory Amplitudes the On-Shell
  Way: Scalar and Vector Couplings to Gluons}},
  \href{https://doi.org/10.1007/JHEP02(2019)165}{\emph{JHEP} {\bfseries 02}
  (2019) 165} [\href{https://arxiv.org/abs/1809.09644}{{\ttfamily
  1809.09644}}].

\bibitem{Christensen:2018zcq}
N.~Christensen and B.~Field, \emph{{Constructive standard model}},
  \href{https://doi.org/10.1103/PhysRevD.98.016014}{\emph{Phys. Rev. D}
  {\bfseries 98} (2018) 016014}
  [\href{https://arxiv.org/abs/1802.00448}{{\ttfamily 1802.00448}}].

\bibitem{Durieux:2019siw}
G.~Durieux and C.~S. Machado, \emph{{Enumerating higher-dimensional operators
  with on-shell amplitudes}},
  \href{https://doi.org/10.1103/PhysRevD.101.095021}{\emph{Phys. Rev. D}
  {\bfseries 101} (2020) 095021}
  [\href{https://arxiv.org/abs/1912.08827}{{\ttfamily 1912.08827}}].

\bibitem{Dong:2022mcv}
Z.-Y. Dong, T.~Ma, J.~Shu and Y.-H. Zheng, \emph{{Constructing generic
  effective field theory for all masses and spins}},
  \href{https://doi.org/10.1103/PhysRevD.106.116010}{\emph{Phys. Rev. D}
  {\bfseries 106} (2022) 116010}
  [\href{https://arxiv.org/abs/2202.08350}{{\ttfamily 2202.08350}}].

\bibitem{Li:2022tec}
H.-L. Li, Z.~Ren, M.-L. Xiao, J.-H. Yu and Y.-H. Zheng, \emph{{Operators for
  generic effective field theory at any dimension: on-shell amplitude basis
  construction}}, \href{https://doi.org/10.1007/JHEP04(2022)140}{\emph{JHEP}
  {\bfseries 04} (2022) 140}
  [\href{https://arxiv.org/abs/2201.04639}{{\ttfamily 2201.04639}}].

\bibitem{Balkin:2021dko}
R.~Balkin, G.~Durieux, T.~Kitahara, Y.~Shadmi and Y.~Weiss, \emph{{On-shell
  Higgsing for EFTs}},
  \href{https://doi.org/10.1007/JHEP03(2022)129}{\emph{JHEP} {\bfseries 03}
  (2022) 129} [\href{https://arxiv.org/abs/2112.09688}{{\ttfamily
  2112.09688}}].

\bibitem{2103.16362}
G.~F.~S. Alves, E.~Bertuzzo and G.~M. Salla, \emph{{On-shell approach to
  neutrino oscillations}},
  \href{https://doi.org/10.1103/PhysRevD.106.036028}{\emph{Phys. Rev. D}
  {\bfseries 106} (2022) 036028}
  [\href{https://arxiv.org/abs/2103.16362}{{\ttfamily 2103.16362}}].

\bibitem{1712.02346}
B.~Schroer, \emph{{The role of positivity and causality in interactions
  involving higher spin}},
  \href{https://doi.org/10.1016/j.nuclphysb.2019.02.007}{\emph{Nucl. Phys. B}
  {\bfseries 941} (2019) 91}
  [\href{https://arxiv.org/abs/1712.02346}{{\ttfamily 1712.02346}}].

\bibitem{1811.01952}
N.~Afkhami-Jeddi, S.~Kundu and A.~Tajdini, \emph{{A Bound on Massive Higher
  Spin Particles}}, \href{https://doi.org/10.1007/JHEP04(2019)056}{\emph{JHEP}
  {\bfseries 04} (2019) 056}
  [\href{https://arxiv.org/abs/1811.01952}{{\ttfamily 1811.01952}}].

\bibitem{2010.02224}
J.~C. Criado, N.~Koivunen, M.~Raidal and H.~Veerm\"ae, \emph{{Dark matter of
  any spin -- an effective field theory and applications}},
  \href{https://doi.org/10.1103/PhysRevD.102.125031}{\emph{Phys. Rev. D}
  {\bfseries 102} (2020) 125031}
  [\href{https://arxiv.org/abs/2010.02224}{{\ttfamily 2010.02224}}].

\bibitem{2011.10058}
T.~Trott, \emph{{Causality, unitarity and symmetry in effective field theory}},
  \href{https://doi.org/10.1007/JHEP07(2021)143}{\emph{JHEP} {\bfseries 07}
  (2021) 143} [\href{https://arxiv.org/abs/2011.10058}{{\ttfamily
  2011.10058}}].

\bibitem{Giunti:1053706}
C.~Giunti and K.~C. Wook, \emph{{Fundamentals of Neutrino Physics and
  Astrophysics}}. Oxford Univ., Oxford, 2007,
  \href{https://doi.org/10.1093/acprof:oso/9780198508717.001.0001}{10.1093/acprof:oso/9780198508717.001.0001}.

\bibitem{barger2012physics}
V.~Barger, D.~Marfatia and K.~Whisnant, \emph{The Physics of Neutrinos}.
  Princeton University Press, 2012.

\bibitem{ParticleDataGroup:2020ssz}
{\scshape Particle Data Group} collaboration, \emph{{Review of Particle
  Physics}}, \href{https://doi.org/10.1093/ptep/ptaa104}{\emph{PTEP} {\bfseries
  2020} (2020) 083C01}.

\bibitem{Wolfenstein:1977ue}
L.~Wolfenstein, \emph{{Neutrino Oscillations in Matter}},
  \href{https://doi.org/10.1103/PhysRevD.17.2369}{\emph{Phys. Rev. D}
  {\bfseries 17} (1978) 2369}.

\bibitem{Mikheyev:1985zog}
S.~P. Mikheyev and A.~Y. Smirnov, \emph{{Resonance Amplification of
  Oscillations in Matter and Spectroscopy of Solar Neutrinos}}, {\emph{Sov. J.
  Nucl. Phys.} {\bfseries 42} (1985) 913}.

\bibitem{Barger:1980tf}
V.~D. Barger, K.~Whisnant, S.~Pakvasa and R.~J.~N. Phillips, \emph{{Matter
  Effects on Three-Neutrino Oscillations}},
  \href{https://doi.org/10.1103/PhysRevD.22.2718}{\emph{Phys. Rev. D}
  {\bfseries 22} (1980) 2718}.

\bibitem{astro-ph/0003365}
W.~Hu, R.~Barkana and A.~Gruzinov, \emph{{Cold and fuzzy dark matter}},
  \href{https://doi.org/10.1103/PhysRevLett.85.1158}{\emph{Phys. Rev. Lett.}
  {\bfseries 85} (2000) 1158}
  [\href{https://arxiv.org/abs/astro-ph/0003365}{{\ttfamily
  astro-ph/0003365}}].

\bibitem{Ferreira:2020fam}
E.~G.~M. Ferreira, \emph{{Ultra-light dark matter}},
  \href{https://doi.org/10.1007/s00159-021-00135-6}{\emph{Astron. Astrophys.
  Rev.} {\bfseries 29} (2021) 7}
  [\href{https://arxiv.org/abs/2005.03254}{{\ttfamily 2005.03254}}].

\bibitem{Arguelles:2015dca}
C.~A. Arg\"uelles, T.~Katori and J.~Salvado, \emph{{New Physics in
  Astrophysical Neutrino Flavor}},
  \href{https://doi.org/10.1103/PhysRevLett.115.161303}{\emph{Phys. Rev. Lett.}
  {\bfseries 115} (2015) 161303}
  [\href{https://arxiv.org/abs/1506.02043}{{\ttfamily 1506.02043}}].

\bibitem{Arguelles:2017atb}
C.~A. Arg\"uelles, A.~Kheirandish and A.~C. Vincent, \emph{{Imaging Galactic
  Dark Matter with High-Energy Cosmic Neutrinos}},
  \href{https://doi.org/10.1103/PhysRevLett.119.201801}{\emph{Phys. Rev. Lett.}
  {\bfseries 119} (2017) 201801}
  [\href{https://arxiv.org/abs/1703.00451}{{\ttfamily 1703.00451}}].

\bibitem{1205.5809}
L.~G. van~den Aarssen, T.~Bringmann and C.~Pfrommer, \emph{{Is dark matter with
  long-range interactions a solution to all small-scale problems of
  \textbackslash{}Lambda CDM cosmology?}},
  \href{https://doi.org/10.1103/PhysRevLett.109.231301}{\emph{Phys. Rev. Lett.}
  {\bfseries 109} (2012) 231301}
  [\href{https://arxiv.org/abs/1205.5809}{{\ttfamily 1205.5809}}].

\bibitem{Stodolsky:1986dx}
L.~Stodolsky, \emph{{On the Treatment of Neutrino Oscillations in a Thermal
  Environment}}, \href{https://doi.org/10.1103/PhysRevD.36.2273}{\emph{Phys.
  Rev. D} {\bfseries 36} (1987) 2273}.

\bibitem{Davoudiasl:2018hjw}
H.~Davoudiasl, G.~Mohlabeng and M.~Sullivan, \emph{{Galactic Dark Matter
  Population as the Source of Neutrino Masses}},
  \href{https://doi.org/10.1103/PhysRevD.98.021301}{\emph{Phys. Rev. D}
  {\bfseries 98} (2018) 021301}
  [\href{https://arxiv.org/abs/1803.00012}{{\ttfamily 1803.00012}}].

\end{thebibliography}\endgroup

\end{document}